\documentclass[namedreferences]{SolarPhysics}
\usepackage[optionalrh]{spr-sola-addons} 
\usepackage{graphicx}        
\usepackage{color}           
\usepackage{url}             

\def\hyph{-\penalty0\hskip0pt\relax}

\newcommand{\aap}{    {\it Astron. Astrophys.}}

\newcommand{\apj}{    {\it Astrophys. J.}}

\newcommand{\jgr}{    {\it J. Geophys. Res.}}

\newcommand{\solphys}{{\it Solar Phys.}}

\begin{document}

\begin{article}

\begin{opening}

\title{Uncertainties in Solar Synoptic Magnetic Flux Maps}
\author{L.~\surname{Bertello}$^{1}$\sep
        A.A.~\surname{Pevtsov}$^{2}$\sep
        G.J.D.~\surname{Petrie}$^{1}$\sep
        D.~\surname{Keys}$^{3}$
       }
\runningauthor{Bertello, et al.}
\runningtitle{Solar Maps}
\institute{$^{1}$National Solar Observatory, Tucson, AZ 85719, USA
   email: \url{bertello@nso.edu} email: \url{petrie@noao.edu} \\
   $^{2}$National Solar Observatory, Sunspot, NM 88349, USA
   email: \url{apevtsov@nso.edu} \\
   $^{3}$University of Arizona, Tucson, AZ 85719, USA
   email: \url{dmkeys@gmail.com}
             }

\begin{abstract}
Magnetic flux synoptic charts are critical for reliable modeling of the 
corona and heliosphere.
Until now, however, these charts were provided without any estimate of uncertainties. 
The uncertainties are due to instrumental noise in the measurements and to the
spatial variance of the magnetic flux distribution that contributes to each bin in the synoptic chart.
We describe here a simple method to compute synoptic magnetic flux maps and their corresponding magnetic
flux spatial variance charts that can be used to estimate the uncertainty in the results of coronal models. 
We have tested this approach by computing a potential\hyph field\hyph source\hyph surface model of the 
coronal field for
a Monte Carlo simulation of Carrington synoptic magnetic flux maps generated from the variance map. 
We show that these uncertainties affect both the locations of source-surface neutral lines
and the distributions of coronal holes in the models.
\end{abstract}
\end{opening}
\section{Introduction}

Synoptic charts are routinely used to display the distribution of various physical quantities over the
entire solar surface. Such charts are indispensable in representing global/large scale properties of
solar magnetic fields in the photosphere and corona including location of active regions and
complexes of activity (active longitudes or activity nests), coronal holes, chromospheric
filaments, and the heliospheric neutral sheet. Synoptic charts of the photospheric magnetic field are
currently used as the input for all coronal and solar wind models. The charts are created by merging
together a series of full disk observations spanning at least a full Carrington rotation. Traditional
synoptic charts are created in several major steps. First, the individual full disk images are remapped
into heliographic coordinates. In case of the photospheric line\hyph of\hyph sight magnetograms an
additional assumption is made that the magnetic field is radial. Second, it is assumed that the solar
surface rotates as a solid body with the rotation rate of $\sim$27.27 days, and the individual remapped
charts belonging to the same Carrington rotation are combined together to form a continuous synoptic
chart. In the process, the parts of remapped charts with overlapping heliographic coordinates are
averaged. 

Due to the non‐rigid rotation, the evolution of solar features and their proper motions, the
averaging tends to smear the features in the synoptic chart as additional remapped images are
added. To minimize this spatial smearing and emphasize the contribution of portion of the solar image
closest to the solar central meridian, NSO (National Solar Observatory) individual remapped images are
typically weighted by a  $\cos ^4 \lambda$ prior to averaging, where $\lambda$ is the central meridian distance. In
addition, to compensate for changes in statistics (number of pixels contributing to each larger pixel
on remapped image), an additional weighting function $\cos \lambda \sin \phi$ (where $\phi$ is
latitude) was applied in early synoptic charts \cite{1980STIN...8121003H}. Later changes to classic synoptic
maps included a latitude dependent “blending” of observed pixels smoothed by a running
Gaussian function to achieve a constant signal\hyph to\hyph noise ratio throughout the chart, and the creation of
dynamic synoptic charts (e.g. \opencite{1998ASPC..140..155H}). The dynamic synoptic charts are updated
by new observations and, thus, represent the evolution of solar features unlike classical synoptic
charts for which each longitudinal strip corresponded to a fixed in time “snap\hyph shot” of solar surface
(usually, when this range of longitudes was near to the solar central meridian). To better represent the
distribution of solar features in between the observational updates, \inlinecite{2000SoPh..195..247W}
developed a concept of evolving synoptic charts, which included the evolution of solar features due
to an average differential rotation profile, meridional flow, supergranular diffusion, and random emergence
of small scale (background) flux elements. 

The Air Force Data Assimilative Photospheric flux Transport (ADAPT) model expands the
\inlinecite{2000SoPh..195..247W} approach.
It includes the Los Alamos National Laboratory (LANL) data assimilation code, which 
uses multiple realizations to account for various model parameters and their uncertainties. 
Observations are incorporated into the model, pixel\hyph by\hyph pixel, by summing the model and observed pixel 
values with weights calculated using a modified least-squares combination of the observational and 
model uncertainties \cite{2010AIPC.1216..343A}. The ADAPT model is used to generate an input of
solar magnetic field maps to models of the background solar wind (Wang\hyph Sheeley\hyph Arge or WSA model) and the
global corona during solar total eclipses (e.g., http://www.nso.edu/node/136).

As for any other measurements, the full disk magnetic observations that form the basis for the synoptic charts are
subject to noise in the measured parameters. The noise could depend on the instrument characteristics
(e.g., type of detector, telescope aperture, pixel size etc, observing conditions (e.g., atmospheric
seeing for ground-based instruments or disturbances in orbital parameters for space‐borne platforms)
and other similar factors. Since the synoptic charts are constructed by averaging a contribution of
individual images with smaller pixels into maps with larger pixels, there are additional statistical
uncertainties. Until now, however, the synoptic maps were produced without corresponding
uncertainties and were essentially treated by modelers as noise\hyph free input. 

Here we present the first
attempt to evaluate the uncertainties of the synoptic charts. As a first step, we concentrate on
statistical uncertainties arising from distribution of image pixels contributing to each pixel in the
synoptic chart. In Section 2, we describe the method to compute these uncertainties and show the
resulting maps of errors. In Section 3, we employ the error maps to create an ensemble of input
synoptic charts within these uncertainties, and we use these charts as input for a Potential Field
Source Surface (PFSS) model (\opencite{1969SoPh....9..131A}, \opencite{1969SoPh....6..442S}).
We show that the level on uncertainties in the synoptic maps may
result in a noticeable change in the projection of the heliospheric neutral line onto the photosphere and
it also affects the position and boundaries of the photospheric footpoints of areas of open magnetic
field (coronal holes). In Section 4 we summarize these findings.

While for this study we used data provided by the Synoptic Optical Long-term Investigations of the 
Sun-Vector SpectroMagnetograph (SOLIS/VSM, \opencite{2011SPIE.8148E...8B} and reference therein), our technique is 
equally applicable to magnetograms produced by other instruments.

\section{Data Analysis}

We describe here a new method to produce remapped heliographic magnetograms and magnetic flux density synoptic charts
from a set of individual magnetogram images. Typically, the heliographic resolution
of either a heliographic magnetogram or a synoptic chart is much lower than the spatial resolution of a single
magnetogram image. Currently, a widely adopted heliographic resolution is $1^{\circ}\times 1^{\circ}$
in latitude and longitude. This resolution implies that, in general, several pixels 
contribute to each heliographic bin. This number, however, varies with the distance from the solar disk center.
Therefore, the distribution of contributing pixels allows not only to compute a weighted mean flux density
for those bins, but also to estimate an uncertainty of this value based on the spatial variance of
this distribution. This procedure is discussed in more detail in the next section.

\subsection{Remapped Heliographic Magnetograms}

\begin{figure}
\begin{center}
\includegraphics[width=1.0\textwidth]{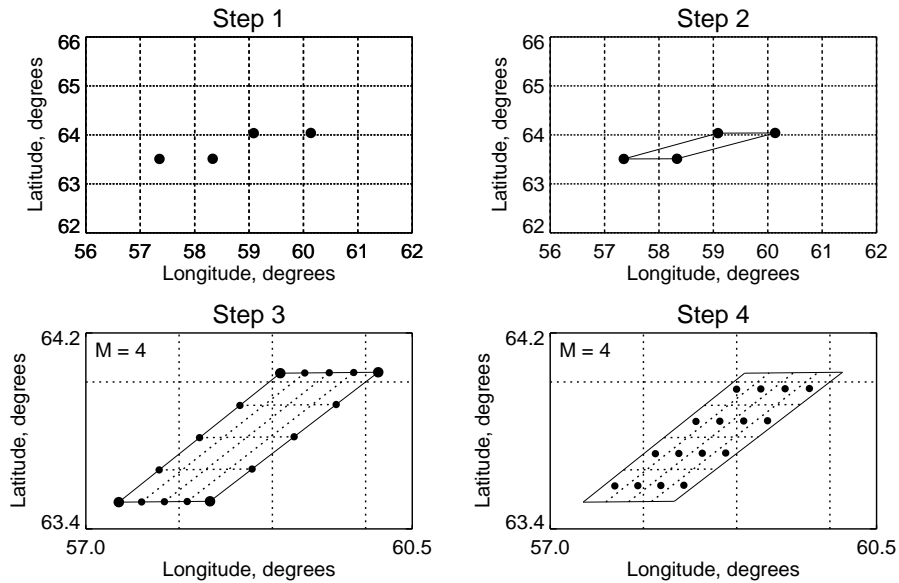}
\caption{Transformation of the four corners of a idealized solar image pixel into heliographic coordinates,
using Eq. \ref{heleqn} (Step 1). Steps 2-4 illustrate the procedure described in Section 2.1 to compute the 
fraction of the magnetic flux density contained in a pixel that contributes
to a particular heliographic bin.}
\label{method}
\end{center}
\end{figure}

To compute the weighted mean flux density in each bin of a heliographic magnetogram 
we follow a 4-step procedure, illustrated in Figure \ref{method}.

The first step consists of transforming the coordinates of the
four corners of a pixel in the magnetogram image into Stonyhurst heliographic longitudes $(L)$ and latitudes 
$(B)$ on the solar disk \cite{2006A&A...449..791T}. If $(x,y)$ are the Cartesian 
coordinates in pixel units of a point in circular magnetogram image relative to the image center
and the position angle between the geocentric North pole and the solar rotational North pole is zero ($P$=0),
the formulas for the computation of the Stonyhurst heliographic coordinates $(L,B)$ are as 
follow (the {\it Astronomical Almanac}):

\begin{eqnarray}
\sin(B) & = & \sin(B_0)\cos(\rho) + \cos(B_0)\sin(\rho)\cos(\theta) \nonumber \\
\sin(L) & = & -\sin(\rho)\sin(\theta)/\cos(B),
\label{heleqn}
\end{eqnarray}

where $\rho = \arcsin(r) - S\cdot r$ is the heliocentric angular distance of the point on the solar surface from the 
center of the Sun's disk, $r = \sqrt{x^2+y^2}/R_{\circ}$, $\theta$ = arg$(x,y)$
\footnote{The function arg is defined such that arg$(x,y) = \arctan(y/x) \in [-180^{\circ},180^{\circ}]$,
and thus resolves the ambiguity between which quadrant the result should lie in.}, $R_{\circ}$ is the solar
radius in pixels, $S$ is the angular semi-diameter of the Sun, and $B_0$ is the Stonyhurst heliographic latitude
of the observer. The top left plot of Figure \ref{method} shows this transformation for the four corners of
a typical off-center pixel. For this illustration we used an ideal $512 \times 512$ image of a solar disk centered
at (256,256) with a radius of 250 pixels, $B_0 = 0$, and semi-diameter $S = 959.63$ arcsec. 

Next, we connect the four points with segments as shown in the top right plot of Figure \ref{method}.
Strictly speaking this is an approximation, since the curve connecting each pair of points is slightly
different than a straight line. However, the difference is negligible because the radius of the Sun is much 
larger than the pixel size, so that to a pixel the radius of curvature is effectively infinite.
The resulting shape is a quadrilateral covering, in some cases, more than a single bin in the heliographic
grid. 

In order to compute the fraction $w$ of the magnetic flux density $B$ contained in a pixel that contributes
to a particular heliographic bin, 
we divide each side of the quadrilateral into 
$M$ equal parts and connect the opposite points, as shown in left bottom panel of
Figure \ref{method} for the case $M = 4$. 
Then, we use the coordinates of the center of each grid point (bottom
right panel of Figure \ref{method}) to compute $w$:
If $n$ is the number of grid points that fall into the heliographic bin, the contribution of this
pixel to the magnetic flux density value of that bin is given by $(n/M^2)B = wB$.
As an example, for the case illustrated in Figure \ref{method} and for the heliographic bin centered at 
$(58.5^{\circ},63.5^{\circ})$, the weight $w$ would be given by 9/16 = 0.5625.
When $M$ is sufficiently high, the difference in the
areas of each grid section becomes negligible and the method is more accurate.
For the SOLIS/VSM data
used in this analysis, we found that a segmentation higher than $M$ = 10 does not
improve the accuracy of the calculations. A comparison with heliographic remapped magnetograms produced
by the SOLIS/VSM pipeline reveals that our remapping method is at least as accurate as other standard techniques
in preserving the magnetic flux. 

The generalization of this procedure for 
the remapping of a magnetogram image into heliographic coordinates is straightforward. If $N_j$
is the number of pixels with values $\{B_i: i=1,\cdots,N_j\}$ and weights
$\{w_i: i=1,\cdots,N_j\}$ that contribute to the heliographic bin $j$, the weighted average flux density
$B_j$ of that bin is given by:

\begin{equation}
B_j = \sum_{i}^{N_j} w_i B_i/W_j,
\label{bj}
\end{equation}

where $W_j = \sum_{i}^{N_j} w_{i}$. 
An unbiased estimator of a weighted population variance can be calculated using the formula:

\begin{equation}
\sigma^2_j = \frac{W_j}{W^2_j - W^{\prime}_j}\sum_{i}^{N_j} w_{i}(B_{i} - B_j)^2,
\label{sj}
\end{equation}

where $W^{\prime}_j = \sum_{i}^{N_j} w^2_{i}$. The standard deviation is simply the square root of the variance above.
Another useful quantity, for the discussion that follows, is:  

\begin{equation}
|B|^2_j = \sum_{i}^{N_j} w_i B^2_i/W_j
\end{equation}

\begin{figure}
\begin{center}
\includegraphics[width=1.0\textwidth]{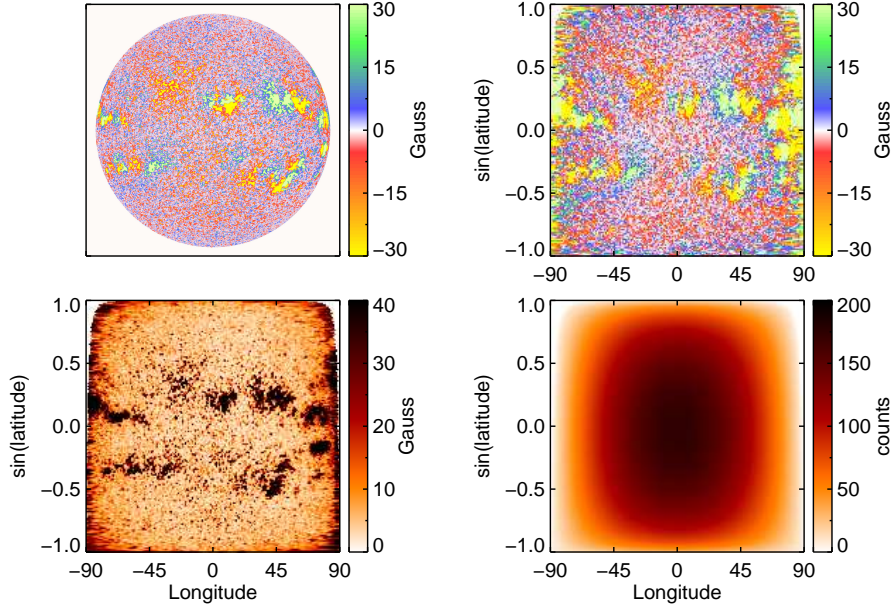}
\caption{
SOLIS/VSM observation taken on May 20, 2013 at 14:15 UT. 
The top left image shows the measured full disk (2048$\times$2048 pixels image) 
longitudinal photospheric magnetic flux density 
in the neutral iron spectral line at 630.15 nm. The top right image is the corresponding heliographic
remapped radial field  magnetogram. The bottom two images show the computed standard deviation map (left) and 
the sum of the weights $W$ (right). The procedure is described in Section 2.1.
}
\label{sample}
\end{center}
\end{figure}

Figure \ref{sample} shows an example of this procedure applied to a 
SOLIS/VSM observation taken on May 20, 2013 at 14:15 UT. Prior of being remapped into heliographic coordinates 
using Eq. \ref{bj}, the original full disk line-of-sight
magnetogram (top left image) is divided by $\cos(\rho)$ to produce a heliographic map of the radial component of
the magnetic flux density (top right image). This transformation, from line-of-sight to radial magnetic field
density, is valid only
under the assumption that the photospheric magnetic field is indeed radial. 
Evidence that the photospheric field is approximately radial was found by 
\inlinecite{1978SoPh...58..225S}, \inlinecite{2009ApJ...699..871P}, and
\inlinecite{2013SoPh..283..195G}.
The standard deviation map (left bottom image) is then computed using Eq. \ref{sj}. 

\subsection{Magnetic Flux Density Synoptic Charts}

Magnetic field observations taken during a single Carrington rotation can be merged together to produce a chart
of the magnetic flux density distribution over the whole solar surface.
If $N$ is the number of magnetograms that contribute to this chart, and $\{N_j: j=1,\cdots,N\}$ is the number of
pixels with flux density and weight values $\{B_{i,j}, w_{i,j}: i = 1,\cdots,N_j\}$ from magnetogram \#j
that contribute to a heliographic bin
$k$ in the synoptic chart, the magnetic flux density $B_k$ of that bin is given by:

\begin{eqnarray}
B_k & = & \left(\sum_{i}^{N_1} w_{i,1}B_{i,1} + \sum_{i}^{N_2} w_{i,2}B_{i,2} +
\cdots + \sum_{i}^{N_N} w_{i,N}B_{i,N}\right)/(W_1 + W_2 + \cdots + W_N) = \nonumber \\
    & = & (W_1B_1 + W_2B_2 + \cdots + W_NB_N)/(W_1 + W_2 + \cdots + W_N) = \nonumber \\
    & = & \sum_{j}^{N} W_{j}B_{j}/  \sum_{j}^{N} W_{j}.
\end{eqnarray}
This shows that $B_k$ can be computed from the values obtained using Eq. \ref{bj}
for the heliographic re-mappped images. 
The number $N$ of magnetograms that contribute to a given synoptic chart and the number $N_j$ of image pixels that go into a 
particular heliographic bin depend on observational conditions. Typically, for SOLIS/VSM  
observations under good weather conditions over a temporal window of 40 days, between 30 and 40 magnetograms are used to 
build a synoptic chart. Each magnetogram contributes with about 30 pixels to individual $1 \times 1$ degrees heliographic 
bins located at low latitudes. This number decreases for bins located at higher latitude bands.
If additional weights are introduced, like $\tilde{w}_j = \cos^4(L_j)$, where $L_j$ is the central 
meridian distance, then the final $B_k$ value will be:

\begin{equation}
B_k =  \sum_{j}^{N} \tilde{w}_j W_{j}B_{j}/  \sum_{j}^{N} \tilde{w}_j W_{j}
\end{equation}

The weight $\tilde{w}$ is typically introduced to ensure that measurements taken when a particular 
Carrington longitude is near the solar central meridian contribute most 
to that Carrington longitude in the synoptic map. 
For the unbiased estimated variance $\sigma_k^2$, associated to $B_k$, we have (see Appendix A):

\begin{equation}
\sigma_k^2  =  \frac{1}{1 -
\left[\sum_{j}^{N} \tilde{w}^2_j W^{\prime}_{j}/\left(\sum_{j}^{N} \tilde{w}_j W_{j}\right)^2\right]} 
\left(\frac{\sum_j^{N} \tilde{w}_j W_j|B|_j^2}{\sum_{j}^{N} \tilde{w}_j W_{j}} - B_k^2\right). 
\label{smap}
\end{equation}

\begin{figure}
\begin{center}
\includegraphics[width=1.0\textwidth]{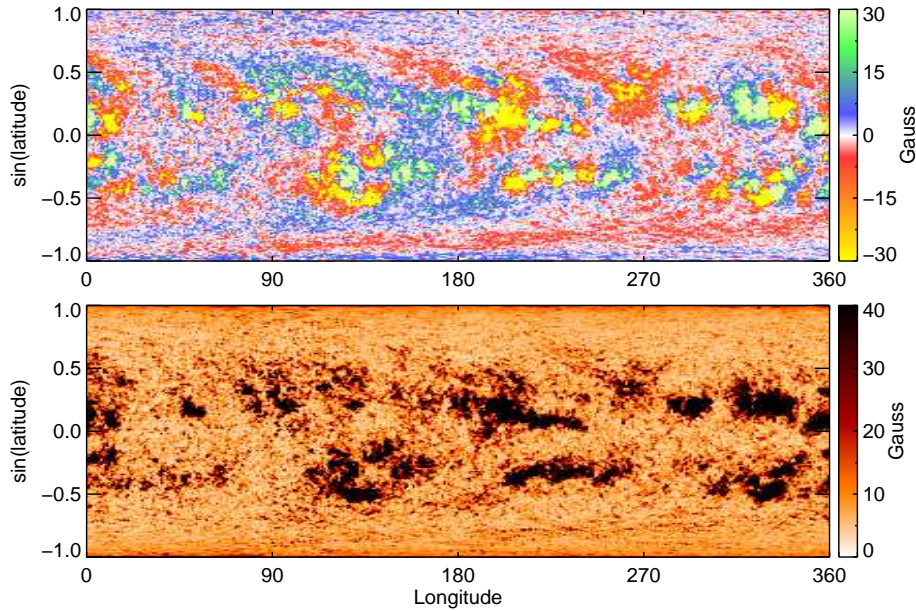}
\caption{
Photospheric magnetic flux density distribution (top) and corresponding standard deviation map (bottom) for CR 2137.
Both charts were computed using Fe I 630.15 nm SOLIS/VSM full disk magnetic observations, following
the procedure described in Section 2.2. The top image has been scaled between $\pm$30 Gauss to better show the distribution of 
the weak magnetic flux density field across the map. The actual flux density distribution, however, covers a much larger 
range of field values - up to several hundred Gauss.
}
\label{cmap}
\end{center}
\end{figure}

The standard deviation synoptic map is simply the square root of Eq. \ref{smap}. Figure \ref{cmap} shows an
example of a photospheric magnetic flux density synoptic chart, and the corresponding standard deviation map.
The charts were computed using Fe I 630.15 nm SOLIS/VSM full disk longitudinal magnetic observations 
covering Carrington rotation (CR) 2137. 
The standard deviation map exhibits several properties. First, the errors show a gradual increase towards polar 
regions (associated with disk center-limb variation in noise). Second, the errors are significantly larger in areas 
associated with strong fields of active regions. This can be related to higher degree of spatial variation in 
magnetic structure as compared with quiet sun areas. Finally, most areas with more uniform fields (e.g, coronal holes)
show the smallest variance.

\begin{figure}
\begin{center}
\includegraphics[width=1.0\textwidth]{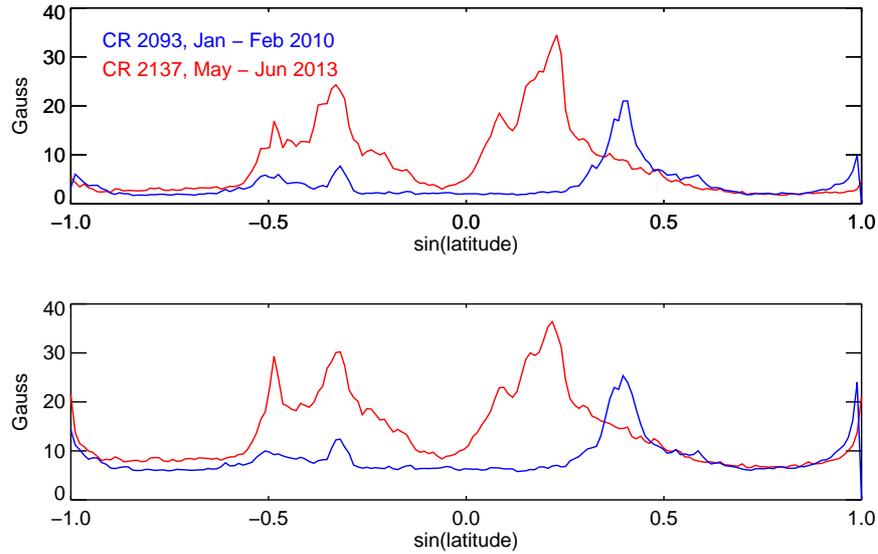}
\caption{Average along the longitudinal direction of the absolute magnetic flux density (top) and standard 
deviation (bottom) maps as a function of latitude, for CR 2093 (blue) and CR 2137 (red). Latitude bands
with high level of activity are also characterized by a larger uncertainty in the magnetic flux density value.
}
\label{elat}
\end{center}
\end{figure}

The high correlation between Carrington error maps and 
Carrington magnetic flux maps is further demonstrated in Figure \ref{elat} where, for each latitude band, we
computed the average of all longitudinal values of the absolute magnetic
flux density (top panel) and the corresponding standard deviation (bottom panel). 
Figure \ref{elat} shows
this comparison during phases of low (CR 2093, in blue) and high (CR 2137, in red) solar magnetic activity. 
At the beginning of 2010
the magnetic activity was predominately concentrated in the North solar hemisphere, in a latitude band centered
just below $30^{\circ}$, while during the current phase of solar maximum the magnetic activity
has developed significantly in both hemispheres. The drastic drop in the mean absolute magnetic flux density
and standard deviation values shown in Figure \ref{elat} near the extreme North polar regions during CR 2093 
is due to the lack of data caused by large negative $B_0$ angle.

\section{Application to Coronal Models}

Since the photospheric synoptic magnetic flux  maps are the primary drivers 
of coronal and heliospheric models, their
accuracy will ultimately affect the diagnostic capabilities of these models.
Our ability to derive global (in solar latitude and longitude) boundary
conditions for these models is hindered because of our limited Earth-based viewpoint.
This affects models in two important ways. First, in order to be able to derive complete
boundary conditions as a function of longitude, we are forced to assume that the
magnetic structure we see from Earth does not appreciably change as
it rotates around to the far side. 
As active regions may significantly evolve on a time scales of one-several days, this is clearly not met.
However, for studies on time-scales of a solar rotation or longer, it is reasonable
to assume that, on average, the unseen decay of active regions will be balanced by
the emergence of new, and as yet unseen active regions.

A second, and potentially more severe limitation for global models is the quality of polar data.
Assuming that the photospheric field is nearly radial, the observed line\hyph of\hyph sight component of
the magnetic field diminishes from equator to pole by a factor $\sim\cos(\rho)$, and
the background noise level rises by the same factor.
Moreover, projection effects also reduce the resolution at progressively higher latitudes.
To further compound this, Earth's ecliptic orbit leads to as much as $\pm 7.25^{\circ}$
offsets between views of the northern and southern polar regions. 
For example, when the solar $B_0$ angle
is +7.25$^{\circ}$, the polar regions near the magnetic pole located in the Sun's northern hemisphere are
observed much better from Earth than are the polar regions around the magnetic pole in the southern
hemisphere. Consequently, for a large fraction of a year, one of the Sun's magnetic
poles is poorly measured.  
The treatment of the Sun's polar fields is difficult but of great importance
in determining the global structure of the corona and the heliospheric magnetic fields (e.g., 
\opencite{1982JGR....8710331H}).
At solar minimum much of the open flux that fills the heliosphere originates in polar
coronal holes. 

Several solutions have been suggested in which the polar field is deduced
using various combinations of spatial and temporal smoothing, interpolation,
and extrapolation from nearby observed regions (e.g. \opencite{2007AAS...210.2405L},
\opencite{2011SoPh..270....9S}). In particular,
\inlinecite{2000JGR...10510465A} have developed a technique to correct polar field measurements
by backward fitting in time during periods when a pole is poorly observed.
For our exploratory investigation, however, it is sufficient to use a simple spatial interpolation approach 
where the value of the magnetic flux density for the unobserved heliographic bins is determined from 
a quintic surface to well\hyph observed fields at polar latitudes. 
The surface fit is similar to the one described in \inlinecite{2011SoPh..270....9S}.

The response of the coronal magnetic field to the photospheric activity patterns can be diagnosed in a simple 
way by calculating extrapolated PFSS models. 
Low in the corona, the magnetic field is sufficiently dominant over the plasma forces that a force-free field 
approximation is generally applicable. Moreover, for large\hyph scale coronal structure the effects of force-free 
electric currents, which are inversely proportional to length scale, may be neglected.
We use the photospheric radial field maps from SOLIS/VSM to fix the radial field component of the model at the 
lower boundary $r=R$, where $R$ is the solar radius. Above some height in the corona, the magnetic field is 
dominated by the thermal pressure and inertial force of the expanding solar wind. To model the effects of the 
solar wind expansion on the field, we introduce an upper boundary at $r=R_s > R$, and force the field to be radial on 
this boundary, following \inlinecite{1969SoPh....6..442S}, \inlinecite{1969SoPh....9..131A} 
and many subsequent authors. 
The usual value for $R_s$ is $R_s=2.5R$ although different choices of $R_s$ lead to more successful 
reconstructions of coronal structure during different phases of the solar cycle (Lee et al.~2011). 
We adopt the standard value $R_s=2.5R$ in this work. With these two boundary conditions, the potential field model 
can be fully determined in the domain $R\le r\le R_s$.

We use the National Center for Atmospheric Research's
MUDPACK\footnote{http://www2.cisl.ucar.edu/resources/legacy/mudpack} package (Adams~1989) to 
solve Laplace's equation numerically in spherical coordinates subject to the above boundary conditions. 
Although Laplace's equation can be solved analytically, and has been so treated for several decades, 
we adopt a finite-difference approach in this paper to avoid some problems associated with the usual approach 
based on spherical harmonics \cite{2011ApJ...732..102T}. 

\begin{figure}
\begin{center}
\includegraphics[width=0.49\textwidth]{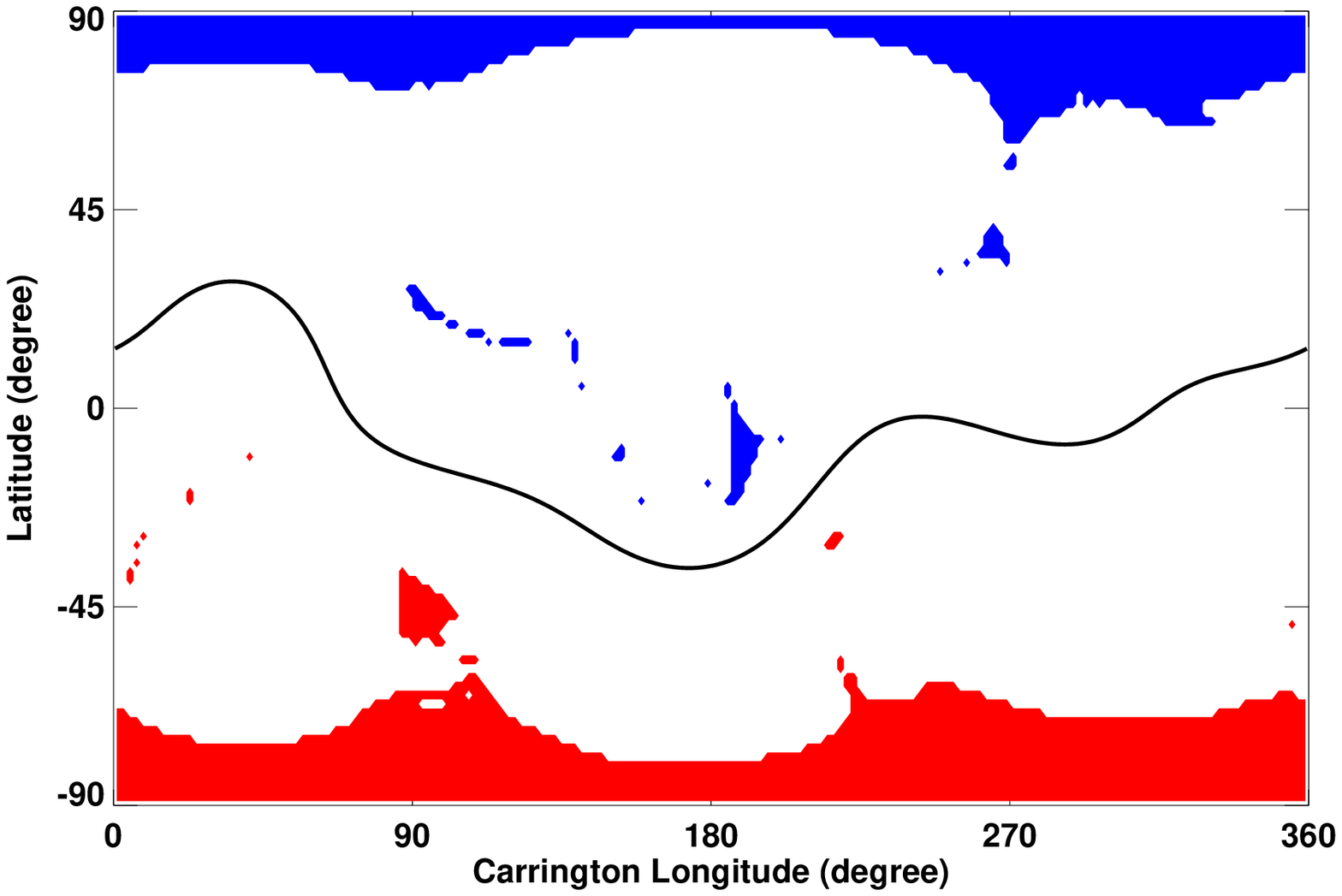}
\includegraphics[width=0.49\textwidth]{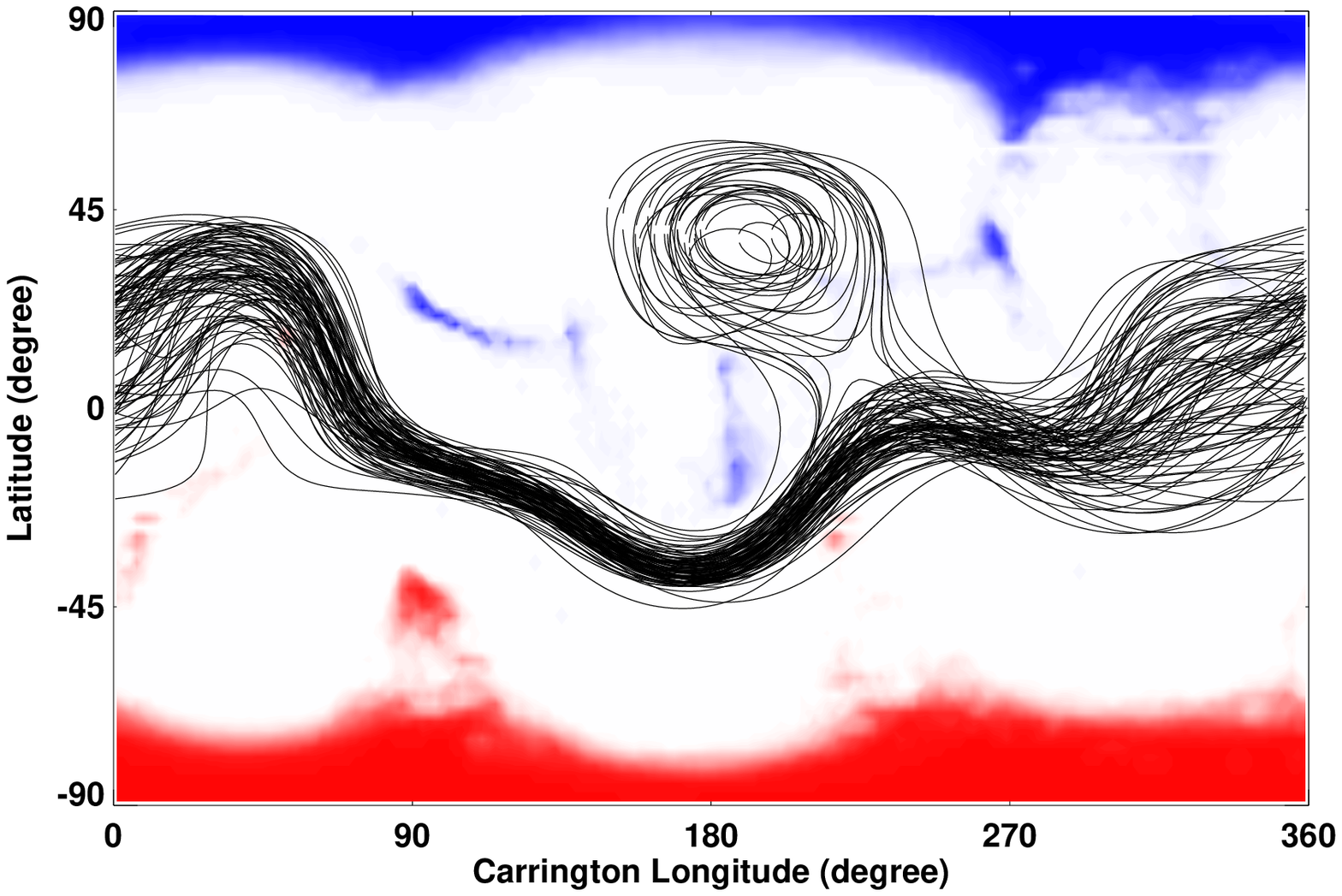}
\caption{
Left, the PFSS neutral line (thin black lines) and positive/negative open field footpoints (red/blue pixels) 
are shown for CR 2104. The open field footpoints correspond to coronal holes and the neutral line represents 
the heliospheric current sheet. Right, the 100 model neutral lines for
a Monte Carlo simulation of Carrington synoptic magnetic flux maps generated from the standard deviation
map are over-plotted. 
Solid red/blue indicates pixels where 100\% of the models have positive/negative open field, 
white represents footpoints where all models have closed field, and stronger/fainter coloring indicates where 
a larger/smaller fraction of the models have open field. 
}
\label{corh1}
\end{center}
\end{figure}

\begin{figure}
\begin{center}
\includegraphics[width=0.49\textwidth]{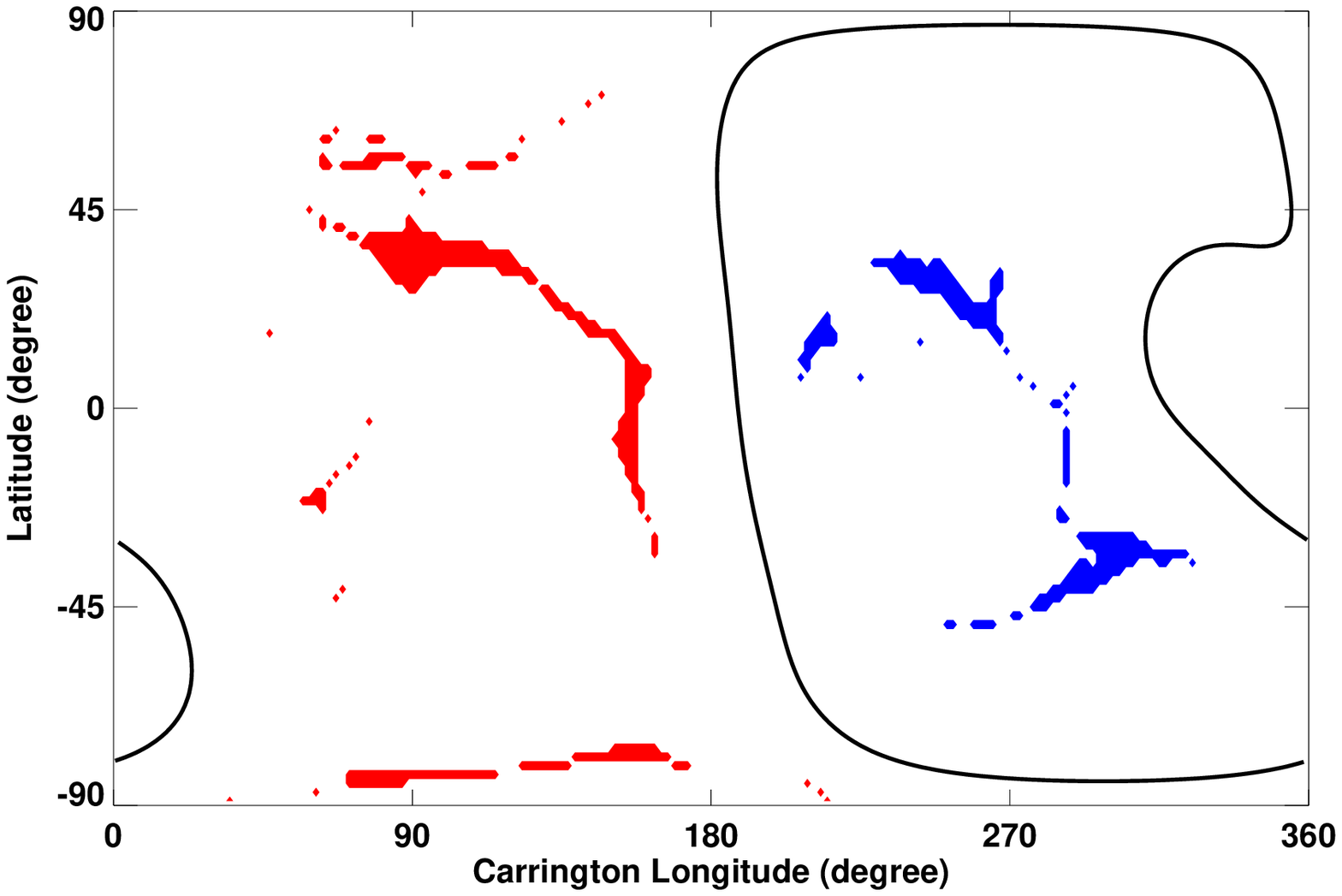}
\includegraphics[width=0.49\textwidth]{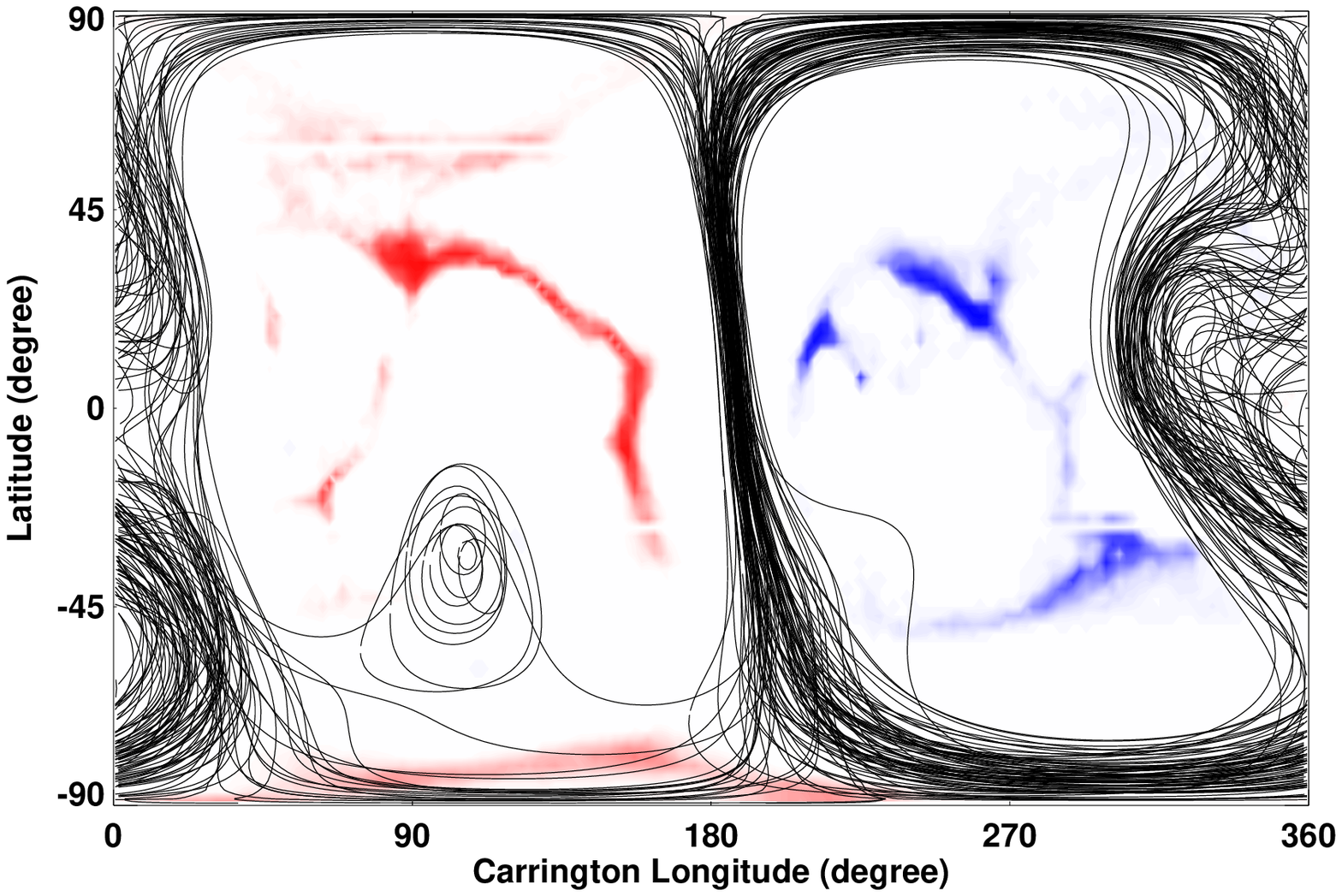}
\caption{
Same as Figure \ref{corh1}, but for CR 2137. 
}
\label{corh2}
\end{center}
\end{figure}

\begin{figure}
\begin{center}
\includegraphics[width=1.00\textwidth]{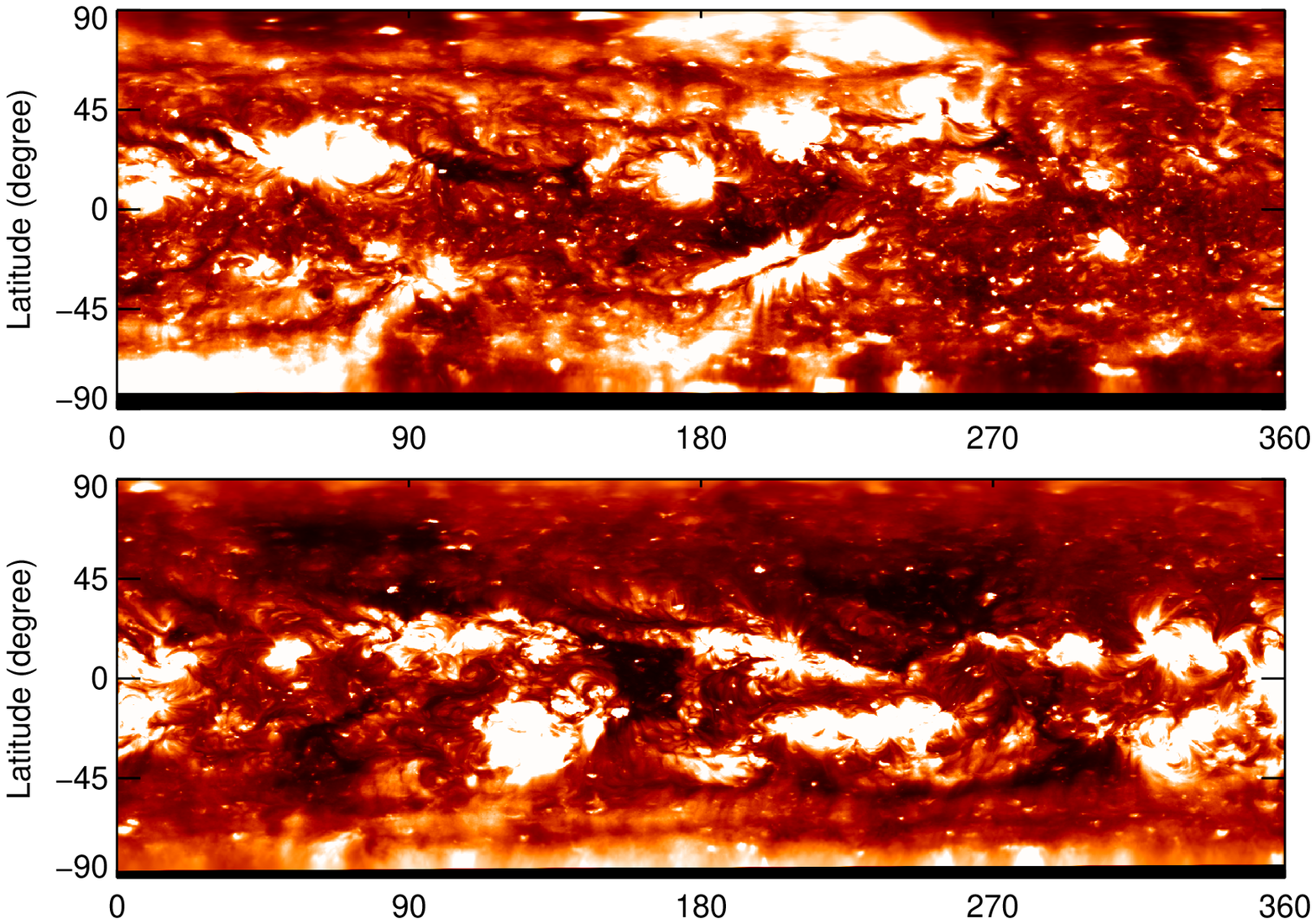}
\caption{
Intensity distribution at $\lambda$195 \AA~from the SECCHI EUV imager instrument for CR 2104 (top) and CR 2137 (bottom).
The images have been enhanced to show the location of coronal holes (black regions).
}
\label{secchi}
\end{center}
\end{figure}

To test how the uncertainties in a synoptic magnetic flux map may affect the calculation of the global
magnetic field of the solar corona we computed a PFSS model of the coronal field for
a Monte Carlo simulation of 100 Carrington synoptic magnetic flux maps generated from the standard deviation  map. 
This test was performed using SOLIS/VSM photospheric longitudinal magnetograms spanning two distinct
Carrington rotation numbers: CR 2104 (November - December 2010) and CR 2137 (May - June 2013). These two CRs
correspond to phases of low and high solar magnetic activity cycle, respectively.
In the simulated synoptic maps of the ensemble, the value of each bin is randomly computed from a normal
distribution with a mean equal to the magnetic flux value of the original bin and a standard deviation
of $\sigma$, with $\sigma$ being the value of the corresponding bin in the standard deviation map.
For each synoptic map of this set we then computed a PFSS model of the coronal field.
Figures \ref{corh1} - \ref{corh2} illustrate the result of this experiment. For comparison we show in
Figure \ref{secchi} measurements at $\lambda$195 \AA~from the SECCHI EUV imager instrument mounted onto the 
NASA STEREO Spacecraft covering the same two Carrington rotation numbers. SECCHI EUV synoptic maps are
available from http://secchi.nrl.navy.mil/synomaps/.

\section{Discussion and Conclusions}

The approach described in this paper is aimed at estimating errors based on
statistical properties of spatial distribution of pixels contributing to heliographic bins. However,
uncertainties due to instrumental noise or other sources can be easily included in the analysis
by properly modifying the equations given in Section 2. 

The CR 2104 global coronal field shown in Figure \ref{corh1} (left), is typical of coronal structure around 
solar minimum: the global dipole is tilted at a small angle with respect to the rotation axis so that the 
positive/negative coronal holes are almost entirely confined to the southern/northern hemisphere and the neutral 
line does not stray more than 40 degrees from the equator. The CR 2137 global coronal field, shown in 
Figure \ref{corh2} (left), is typical of coronal structure around solar maximum: the global dipole is tilted at 
an angle close to 90 degrees with respect to the rotation axis. The positive/negative coronal holes are 
confined to the eastern/western hemisphere and the neutral line encircles the Sun passing close to both poles. 
The comparison between these two coronal maps and the observed coronal holes by the SECCHI EUV imager instrument
(Figure \ref{secchi}) shows a general good agreement between observations and modeling. 
For example, the area of open magnetic field in Figure \ref{corh1} (left) situated at about (-15, 200) degrees 
latitude-longitude corresponds to a position of small coronal hole in Figure \ref{secchi} (upper panel). 
For CR 2137, the area of open field in Figure \ref{corh2} (left) at about (-45, 300) degrees 
overlaps with coronal hole in 
Figure \ref{secchi} (lower panel). 

The Monte Carlo simulations for both rotations (right panel of Figure \ref{corh1} and Figure \ref{corh2})
show significant diversity of neutral line structure. 
In some of the CR 2104 models there are two separate neutral lines, one resembling the neutral line of the original 
model and a secondary compact, closed neutral line centered near 180 degrees longitude, 45 degrees latitude. 
In the original CR 2137 model the neutral line is mostly confined to the western hemisphere, whereas in a significant 
number of the models in the Monte Carlo simulation the neutral lines are shifted into the eastern hemisphere. 
In a subset of the simulations a secondary compact, closed neutral line appears near 90 degrees longitude, 
-45 degrees latitude.
The coronal hole distributions of the Monte Carlo simulations closely resemble the distributions in the original 
models in general. However, the faint coloring of some of the coronal hole structures in the right panels 
of Figures \ref{corh1} and \ref{corh2} indicates that some large coronal hole structures, 
e.g., the most northerly and southerly 
red patches in the eastern hemisphere in Figure \ref{corh2}, are absent from a significant proportion of the models. 

The global coronal structure depends most on the polar fields, which give the overall large-scale axisymmetric structure, 
and the active regions, which perturb the global field into complex 3D configurations. 
Although the polar field errors are lower than those
associated to active regions (see Figure \ref{cmap}), 
they are associated with nearly unipolar fluxes that extend over vast areas, and therefore have major global influence 
over the models. The polar field strengths themselves are typically about 5 Gauss and the relative errors in the polar fields 
are significant. During solar minima the polar fields dominate and during maxima the active regions are more influential. 
These influences can be quantified and compared (\opencite{2013ApJ...768..162P}). 
Therefore, errors in both the polar or active region fields would affect the models, with greatest influence during 
different phases of the cycle: Polar fields contribute most of the errors near solar minimum, while
active regions contribute most of the errors near maximum.
This implies that the errors from the polar fields and active regions have about the same overall influence over the models.
Figures \ref{corh1} and \ref{corh2}, representing near-minimum and near-maximum fields, indicate that the errors from 
the polar fields and active regions have about the same overall influence over the models.

We therefore conclude that the errors have a significant influence on the location and structure of the neutral 
lines and the distribution of the coronal holes.
These initial results indicate that synoptic error maps can play an important role for model 
predictions based on global distribution of magnetic fields on the Sun. In the future, these effects need 
to be included in all global models. In this article, we considered only the effects of statistical 
dispersion of image pixels contributing to the same heliographic pixel of synoptic chart. 
Other (more minor) effects related to instrument and observations (e.g., noise, atmospheric seeing etc) 
will be the subject of a separate future study.

Finally, it is important to notice that while our study clearly shows the significance of introducing uncertainties 
into the computation
of the extrapolated coronal field, several factors may affect the final result. For example, although synoptic charts 
produced by different observatories are very similar, the calibration of the magnetic field measurements and the assumptions 
adopted in constructing the synoptic map would differ from one observatory to another. These differences my include
the treatment of solar differential rotation, evolution of active regions, and the filling of the polar regions. All
these factors will affect not only the final synoptic magnetic chart, but also the corresponding error map. A detailed study of
these differences is beyond the scope of this study.
The purpose of our investigation is to show that taking into account errors in synoptic maps is extremely important, regardless of the
particular adopted methodology and/or instrument.

\appendix

\section{Estimated Variance in Magnetic Flux Synoptic Charts}

Using the notation described in Section 2.2, one can generalize Eq. \ref{sj} to compute the variance
$\sigma_k^2$ of a heliographic bin $k$ in the synoptic chart given the contribution of $N$ individual
observations. That is,

\begin{eqnarray*}
\sigma_k^2  & = &\frac{\tilde{w}_1 W_1+\cdots+\tilde{w}_N W_N}{\left(\tilde{w}_1 W_1+\cdots+\tilde{w}_N W_N\right)^2
-\left(\tilde{w}^2_1 W^{\prime}_1+\cdots+\tilde{w}^2_N W^{\prime}_N\right)}
   \left[\sum_{i}^{N_1}\tilde{w}_1 w_{i,1}(B_{i,1} - B_k)^2 + \right.\\
        & + & \left. \cdots + \sum_{i}^{N_N}\tilde{w}_N w_{i,N}(B_{i,N} - B_k)^2\right] = \\
  & = & \frac{\sum_{j}^{N} \tilde{w}_j W_{j}}{\left(\sum_{j}^{N} \tilde{w}_j W_{j}\right)^2 -
\sum_{j}^{N} \tilde{w}^2_j W^{\prime}_{j}}\left[\tilde{w}_1W_1(|B|_1^2 + B_k^2 - 2B_1B_k) + 
\cdots + \right. \\
  & + & \left. \tilde{w}_NW_N(|B|_N^2 + B_k^2 - 2B_NB_k)\right] = \\
  & = & \frac{\sum_{j}^{N} \tilde{w}_j W_{j}}{\left(\sum_{j}^{N} \tilde{w}_j W_{j}\right)^2 -
\sum_{j}^{N} \tilde{w}^2_j W^{\prime}_{j}}  \sum_j^{N} \tilde{w}_j W_j(|B|_j^2 + B_k^2 - 2B_jB_k)
\end{eqnarray*}

With some manipulation, the previous formula can be written as:
\begin{equation}
\sigma_k^2  =  \frac{\left(\sum_{j}^{N} \tilde{w}_j W_{j}\right)^2}{\left(\sum_{j}^{N} \tilde{w}_j W_{j}\right)^2 -
\sum_{j}^{N} \tilde{w}^2_j W^{\prime}_{j}} 
\left(\frac{\sum_j^{N} \tilde{w}_j W_j|B|_j^2}{\sum_{j}^{N} \tilde{w}_j W_{j}} - B_k^2\right), 
\end{equation}
or simply,

\begin{equation}
\sigma_k^2  =  \frac{1}{1 -
\left[\sum_{j}^{N} \tilde{w}^2_j W^{\prime}_{j}/\left(\sum_{j}^{N} \tilde{w}_j W_{j}\right)^2\right]} 
\left(\frac{\sum_j^{N} \tilde{w}_j W_j|B|_j^2}{\sum_{j}^{N} \tilde{w}_j W_{j}} - B_k^2\right). 
\end{equation}

The standard deviation is simply the square root of the variance above.

\begin{acks}
The authors acknowledge fruitful discussions with Anna Hughes, Jack Harvey, Janet Luhmann, and Thomas Wentzel. 
This work utilizes SOLIS/VSM data obtained by the NSO Integrated Synoptic Program (NISP), managed by the National 
Solar Observatory, which is operated by the Association of Universities for Research in Astronomy (AURA), Inc. 
under a cooperative agreement with the National Science Foundation.
\end{acks}


\begin{thebibliography}{17}
\ifx \bisbn   \undefined \def \bisbn  #1{ISBN #1}\fi
\ifx \binits  \undefined \def \binits#1{#1}\fi
\ifx \bauthor  \undefined \def \bauthor#1{#1}\fi
\ifx \batitle  \undefined \def \batitle#1{#1}\fi
\ifx \bjtitle  \undefined \def \bjtitle#1{\textit{#1}}\fi
\ifx \bvolume  \undefined \def \bvolume#1{\textbf{#1}}\fi
\ifx \byear  \undefined \def \byear#1{#1}\fi
\ifx \bissue  \undefined \def \bissue#1{#1}\fi
\ifx \bfpage  \undefined \def \bfpage#1{#1}\fi
\ifx \blpage  \undefined \def \blpage #1{#1}\fi
\ifx \burl  \undefined \def \burl#1{\textsf{#1}}\fi
\ifx \href  \undefined \def \href#1#2{\textsf{#2}}\fi
\ifx \doiurl  \undefined \def
  \doiurl#1{\href{http://dx.doi.org/#1}{\textsf{#1}}}\fi
\ifx \betal  \undefined \def \betal{\textit{et al.}}\fi
\ifx \binstitute  \undefined \def \binstitute#1{#1}\fi
\ifx \bctitle  \undefined \def \bctitle#1{#1}\fi
\ifx \beditor  \undefined \def \beditor#1{#1}\fi
\ifx \bpublisher  \undefined \def \bpublisher#1{#1}\fi
\ifx \bbtitle  \undefined \def \bbtitle#1{\textit{#1}}\fi
\ifx \bedition  \undefined \def \bedition#1{#1}\fi
\ifx \bseriesno  \undefined \def \bseriesno#1{\textbf{#1}}\fi
\ifx \blocation  \undefined \def \blocation#1{#1}\fi
\ifx \bsertitle  \undefined \def \bsertitle#1{\textit{#1}}\fi
\ifx \bsnm \undefined \def \bsnm#1{#1}\fi
\ifx \bsuffix \undefined \def \bsuffix#1{#1}\fi
\ifx \bparticle \undefined \def \bparticle#1{#1}\fi
\ifx \barticle \undefined \def \barticle#1{}\fi
\ifx \botherref \undefined \def \botherref#1{}\fi
\ifx \url \undefined \def \url#1{\textsf{#1}}\fi
\ifx \bchapter \undefined \def \bchapter#1{}\fi
\ifx \bbook \undefined \def \bbook#1{}\fi
\ifx \bcomment \undefined \def \bcomment#1{#1}\fi
\ifx \oauthor \undefined \def \oauthor#1{#1}\fi
\ifx \citeauthoryear \undefined \def \citeauthoryear#1{#1}\fi
\def \endbibitem {}
\ifx \bconflocation  \undefined \def \bconflocation#1{#1} \fi

\bibitem[\protect\citeauthoryear{{Altschuler} and
  {Newkirk}}{1969}]{1969SoPh....9..131A}
\begin{barticle}
\bauthor{\bsnm{{Altschuler}}, \binits{M.D.}},
\bauthor{\bsnm{{Newkirk}}, \binits{G.}}:
\byear{1969},
\batitle{{Magnetic Fields and the Structure of the Solar Corona. I: Methods of
  Calculating Coronal Fields}}.
\bjtitle{\solphys}
\bvolume{9},
\bfpage{131}\,--\,\blpage{149}.
doi:\doiurl{10.1007/BF00145734}.
\end{barticle}
\endbibitem

\bibitem[\protect\citeauthoryear{{Arge} and
  {Pizzo}}{2000}]{2000JGR...10510465A}
\begin{barticle}
\bauthor{\bsnm{{Arge}}, \binits{C.N.}},
\bauthor{\bsnm{{Pizzo}}, \binits{V.J.}}:
\byear{2000},
\batitle{{Improvement in the prediction of solar wind conditions using
  near-real time solar magnetic field updates}}.
\bjtitle{\jgr}
\bvolume{105},
\bfpage{10465}\,--\,\blpage{10480}.
doi:\doiurl{10.1029/1999JA000262}.
\end{barticle}
\endbibitem

\bibitem[\protect\citeauthoryear{{Arge}
  \textit{et~al.}}{2010}]{2010AIPC.1216..343A}
\begin{barticle}
\bauthor{\bsnm{{Arge}}, \binits{C.N.}},
\bauthor{\bsnm{{Henney}}, \binits{C.J.}},
\bauthor{\bsnm{{Koller}}, \binits{J.}},
\bauthor{\bsnm{{Compeau}}, \binits{C.R.}},
\bauthor{\bsnm{{Young}}, \binits{S.}},
\bauthor{\bsnm{{MacKenzie}}, \binits{D.}},
\bauthor{\bsnm{{Fay}}, \binits{A.}},
\bauthor{\bsnm{{Harvey}}, \binits{J.W.}}:
\byear{2010},
\batitle{{Air Force Data Assimilative Photospheric Flux Transport (ADAPT)
  Model}}.
\bjtitle{Twelfth International Solar Wind Conference}
\bvolume{1216},
\bfpage{343}\,--\,\blpage{346}.
doi:\doiurl{10.1063/1.3395870}.
\end{barticle}
\endbibitem

\bibitem[\protect\citeauthoryear{{Balasubramaniam} and
  {Pevtsov}}{2011}]{2011SPIE.8148E...8B}
\begin{bchapter}
\bauthor{\bsnm{{Balasubramaniam}}, \binits{K.S.}},
\bauthor{\bsnm{{Pevtsov}}, \binits{A.}}:
\byear{2011},
\bctitle{{Ground-based synoptic instrumentation for solar observations}}.
In: \bbtitle{Society of Photo-Optical Instrumentation Engineers (SPIE)
  Conference Series},
\bsertitle{Society of Photo-Optical Instrumentation Engineers (SPIE) Conference
  Series}
\bseriesno{8148},
\bfpage{814809}.
doi:\doiurl{10.1117/12.892824}.
\end{bchapter}
\endbibitem

\bibitem[\protect\citeauthoryear{{Gosain} and
  {Pevtsov}}{2013}]{2013SoPh..283..195G}
\begin{barticle}
\bauthor{\bsnm{{Gosain}}, \binits{S.}},
\bauthor{\bsnm{{Pevtsov}}, \binits{A.A.}}:
\byear{2013},
\batitle{{Resolving Azimuth Ambiguity Using Vertical Nature of Solar Quiet-Sun
  Magnetic Fields}}.
\bjtitle{\solphys}
\bvolume{283},
\bfpage{195}\,--\,\blpage{205}.
doi:\doiurl{10.1007/s11207-012-0135-1}.
\end{barticle}
\endbibitem

\bibitem[\protect\citeauthoryear{{Harvey} and
  {Worden}}{1998}]{1998ASPC..140..155H}
\begin{bchapter}
\bauthor{\bsnm{{Harvey}}, \binits{J.}},
\bauthor{\bsnm{{Worden}}, \binits{J.}}:
\byear{1998},
\bctitle{{New Types and Uses of Synoptic Maps}}.
In: \beditor{\bsnm{{Balasubramaniam}}, \binits{K.S.}},
\beditor{\bsnm{{Harvey}}, \binits{J.}},
\beditor{\bsnm{{Rabin}}, \binits{D.}} (eds.)
\bbtitle{Synoptic Solar Physics},
\bsertitle{Astronomical Society of the Pacific Conference Series}
\bseriesno{140},
\bfpage{155}.
\end{bchapter}
\endbibitem

\bibitem[\protect\citeauthoryear{{Harvey}
  \textit{et~al.}}{1980}]{1980STIN...8121003H}
\begin{barticle}
\bauthor{\bsnm{{Harvey}}, \binits{J.}},
\bauthor{\bsnm{{Gillespie}}, \binits{B.}},
\bauthor{\bsnm{{Miedaner}}, \binits{P.}},
\bauthor{\bsnm{{Slaughter}}, \binits{C.}}:
\byear{1980},
\batitle{{Synoptic solar magnetic field maps for the interval including
  Carrington Rotation 1601-1680, May 5, 1973 - April 26, 1979}}.
\bjtitle{NASA STI/Recon Technical Report N}
\bvolume{81},
\bfpage{21003}.
\end{barticle}
\endbibitem

\bibitem[\protect\citeauthoryear{{Hoeksema}, {Wilcox}, and
  {Scherrer}}{1982}]{1982JGR....8710331H}
\begin{barticle}
\bauthor{\bsnm{{Hoeksema}}, \binits{J.T.}},
\bauthor{\bsnm{{Wilcox}}, \binits{J.M.}},
\bauthor{\bsnm{{Scherrer}}, \binits{P.H.}}:
\byear{1982},
\batitle{{Structure of the heliospheric current sheet in the early portion of
  sunspot cycle 21}}.
\bjtitle{\jgr}
\bvolume{87},
\bfpage{10331}\,--\,\blpage{10338}.
doi:\doiurl{10.1029/JA087iA12p10331}.
\end{barticle}
\endbibitem

\bibitem[\protect\citeauthoryear{{Liu}
  \textit{et~al.}}{2007}]{2007AAS...210.2405L}
\begin{bchapter}
\bauthor{\bsnm{{Liu}}, \binits{Y.}},
\bauthor{\bsnm{{Hoeksema}}, \binits{J.T.}},
\bauthor{\bsnm{{Zhao}}, \binits{X.}},
\bauthor{\bsnm{{Larson}}, \binits{R.M.}}:
\byear{2007},
\bctitle{{MDI Synoptic Charts of Magnetic Field: Interpolation of Polar
  Fields}}.
In: \bbtitle{American Astronomical Society Meeting Abstracts \#210},
\bsertitle{Bulletin of the American Astronomical Society}
\bseriesno{39},
\bfpage{129}.
\end{bchapter}
\endbibitem

\bibitem[\protect\citeauthoryear{{Petrie}}{2013}]{2013ApJ...768..162P}
\begin{barticle}
\bauthor{\bsnm{{Petrie}}, \binits{G.J.D.}}:
\byear{2013},
\batitle{{Solar Magnetic Activity Cycles, Coronal Potential Field Models and
  Eruption Rates}}.
\bjtitle{\apj}
\bvolume{768},
\bfpage{162}.
doi:\doiurl{10.1088/0004-637X/768/2/162}.
\end{barticle}
\endbibitem

\bibitem[\protect\citeauthoryear{{Petrie} and
  {Patrikeeva}}{2009}]{2009ApJ...699..871P}
\begin{barticle}
\bauthor{\bsnm{{Petrie}}, \binits{G.J.D.}},
\bauthor{\bsnm{{Patrikeeva}}, \binits{I.}}:
\byear{2009},
\batitle{{A Comparative Study of Magnetic Fields in the Solar Photosphere and
  Chromosphere at Equatorial and Polar Latitudes}}.
\bjtitle{\apj}
\bvolume{699},
\bfpage{871}\,--\,\blpage{884}.
doi:\doiurl{10.1088/0004-637X/699/1/871}.
\end{barticle}
\endbibitem

\bibitem[\protect\citeauthoryear{{Schatten}, {Wilcox}, and
  {Ness}}{1969}]{1969SoPh....6..442S}
\begin{barticle}
\bauthor{\bsnm{{Schatten}}, \binits{K.H.}},
\bauthor{\bsnm{{Wilcox}}, \binits{J.M.}},
\bauthor{\bsnm{{Ness}}, \binits{N.F.}}:
\byear{1969},
\batitle{{A model of interplanetary and coronal magnetic fields}}.
\bjtitle{\solphys}
\bvolume{6},
\bfpage{442}\,--\,\blpage{455}.
doi:\doiurl{10.1007/BF00146478}.
\end{barticle}
\endbibitem

\bibitem[\protect\citeauthoryear{{Sun}
  \textit{et~al.}}{2011}]{2011SoPh..270....9S}
\begin{barticle}
\bauthor{\bsnm{{Sun}}, \binits{X.}},
\bauthor{\bsnm{{Liu}}, \binits{Y.}},
\bauthor{\bsnm{{Hoeksema}}, \binits{J.T.}},
\bauthor{\bsnm{{Hayashi}}, \binits{K.}},
\bauthor{\bsnm{{Zhao}}, \binits{X.}}:
\byear{2011},
\batitle{{A New Method for Polar Field Interpolation}}.
\bjtitle{\solphys}
\bvolume{270},
\bfpage{9}\,--\,\blpage{22}.
doi:\doiurl{10.1007/s11207-011-9751-4}.
\end{barticle}
\endbibitem

\bibitem[\protect\citeauthoryear{{Svalgaard}, {Duvall}, and
  {Scherrer}}{1978}]{1978SoPh...58..225S}
\begin{barticle}
\bauthor{\bsnm{{Svalgaard}}, \binits{L.}},
\bauthor{\bsnm{{Duvall}}, \binits{T.L.} \bsuffix{Jr.}},
\bauthor{\bsnm{{Scherrer}}, \binits{P.H.}}:
\byear{1978},
\batitle{{The strength of the sun's polar fields}}.
\bjtitle{\solphys}
\bvolume{58},
\bfpage{225}\,--\,\blpage{239}.
doi:\doiurl{10.1007/BF00157268}.
\end{barticle}
\endbibitem

\bibitem[\protect\citeauthoryear{{Thompson}}{2006}]{2006A&A...449..791T}
\begin{barticle}
\bauthor{\bsnm{{Thompson}}, \binits{W.T.}}:
\byear{2006},
\batitle{{Coordinate systems for solar image data}}.
\bjtitle{\aap}
\bvolume{449},
\bfpage{791}\,--\,\blpage{803}.
doi:\doiurl{10.1051/0004-6361:20054262}.
\end{barticle}
\endbibitem

\bibitem[\protect\citeauthoryear{{T{\'o}th}, {van der Holst}, and
  {Huang}}{2011}]{2011ApJ...732..102T}
\begin{barticle}
\bauthor{\bsnm{{T{\'o}th}}, \binits{G.}},
\bauthor{\bsnm{{van der Holst}}, \binits{B.}},
\bauthor{\bsnm{{Huang}}, \binits{Z.}}:
\byear{2011},
\batitle{{Obtaining Potential Field Solutions with Spherical Harmonics and
  Finite Differences}}.
\bjtitle{\apj}
\bvolume{732},
\bfpage{102}.
doi:\doiurl{10.1088/0004-637X/732/2/102}.
\end{barticle}
\endbibitem

\bibitem[\protect\citeauthoryear{{Worden} and
  {Harvey}}{2000}]{2000SoPh..195..247W}
\begin{barticle}
\bauthor{\bsnm{{Worden}}, \binits{J.}},
\bauthor{\bsnm{{Harvey}}, \binits{J.}}:
\byear{2000},
\batitle{{An Evolving Synoptic Magnetic Flux map and Implications for the
  Distribution of Photospheric Magnetic Flux}}.
\bjtitle{\solphys}
\bvolume{195},
\bfpage{247}\,--\,\blpage{268}.
doi:\doiurl{10.1023/A:1005272502885}.
\end{barticle}
\endbibitem

\end{thebibliography}

\end{article} 
\end{document}